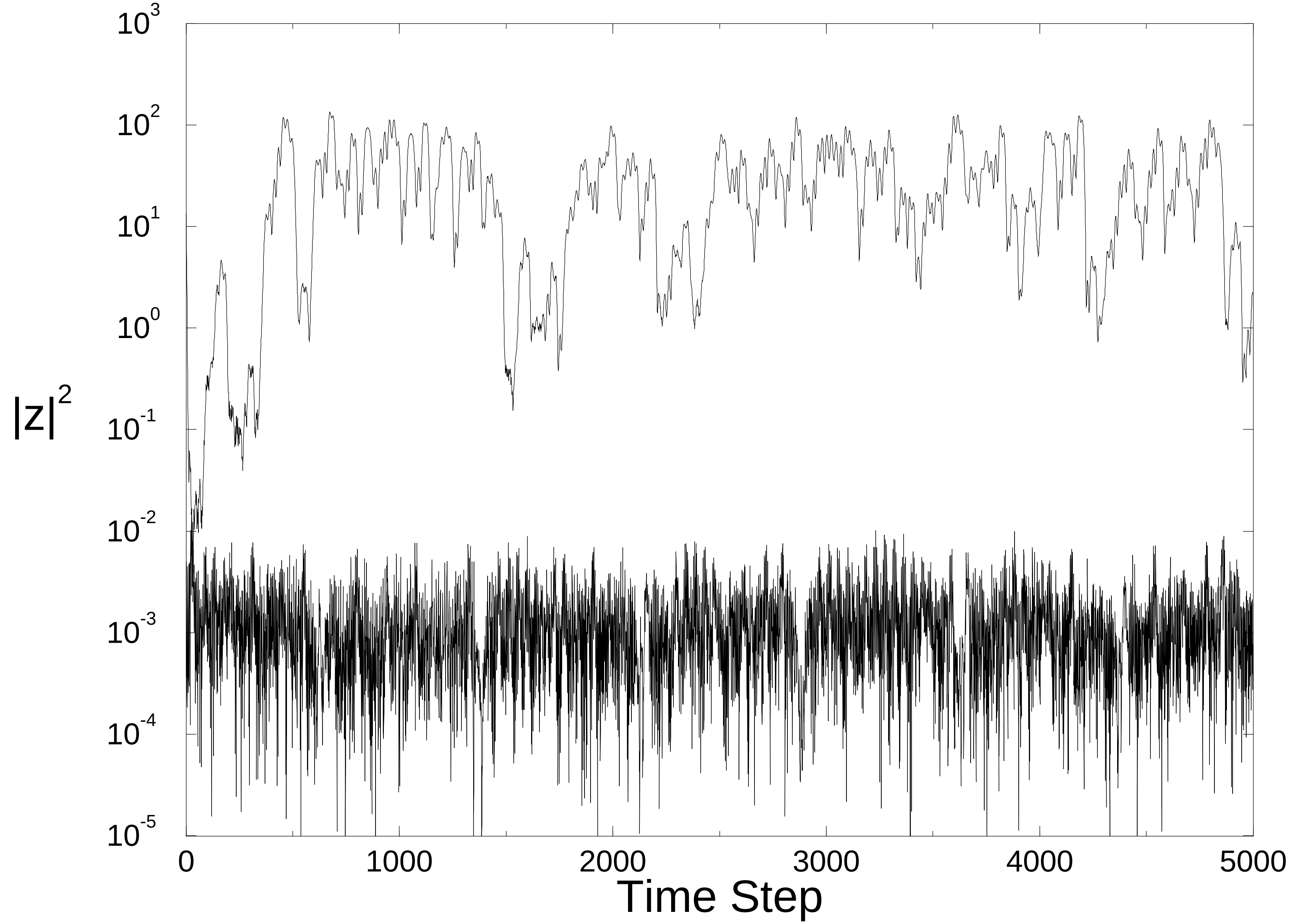

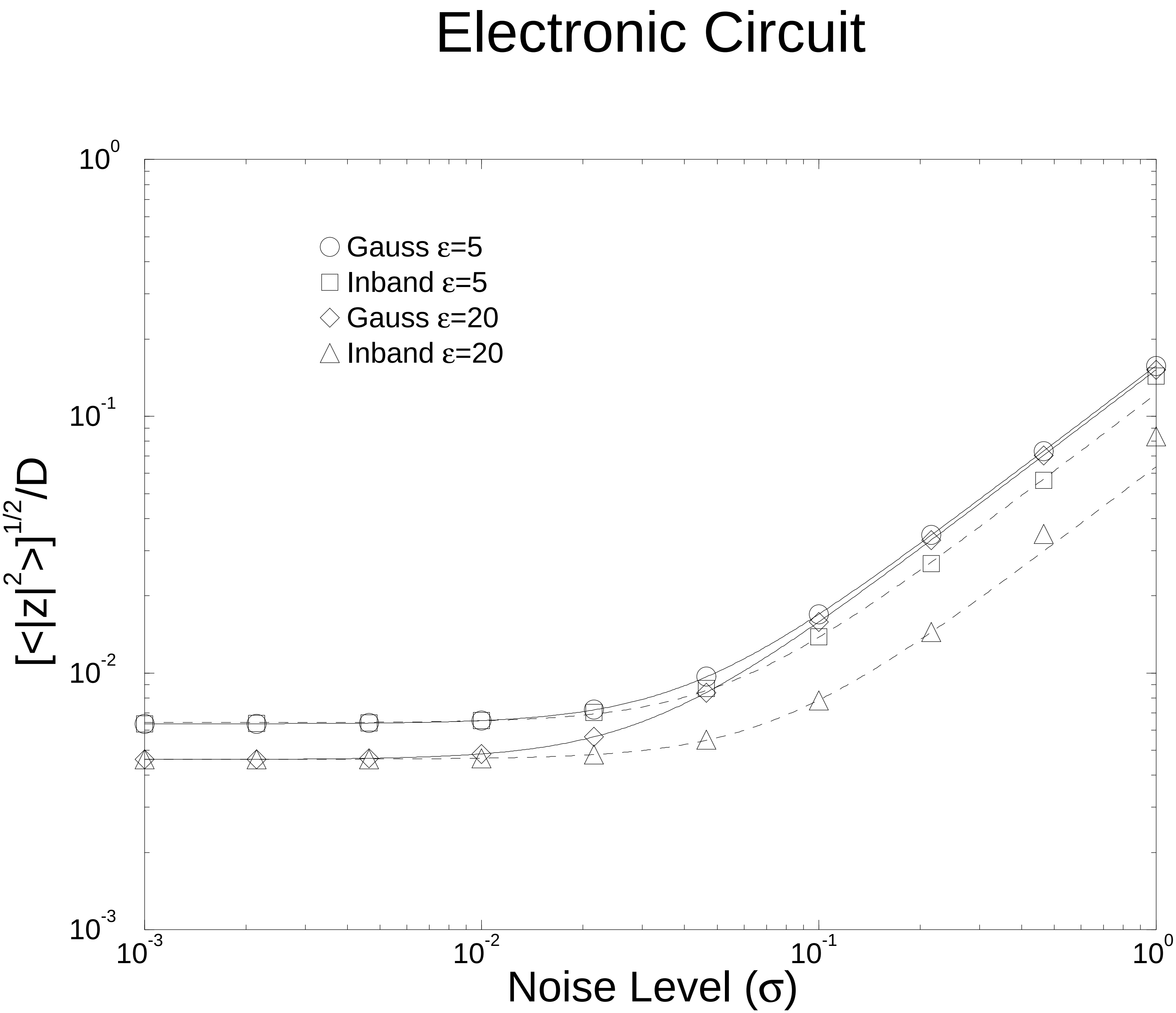

# Vibrating Wire

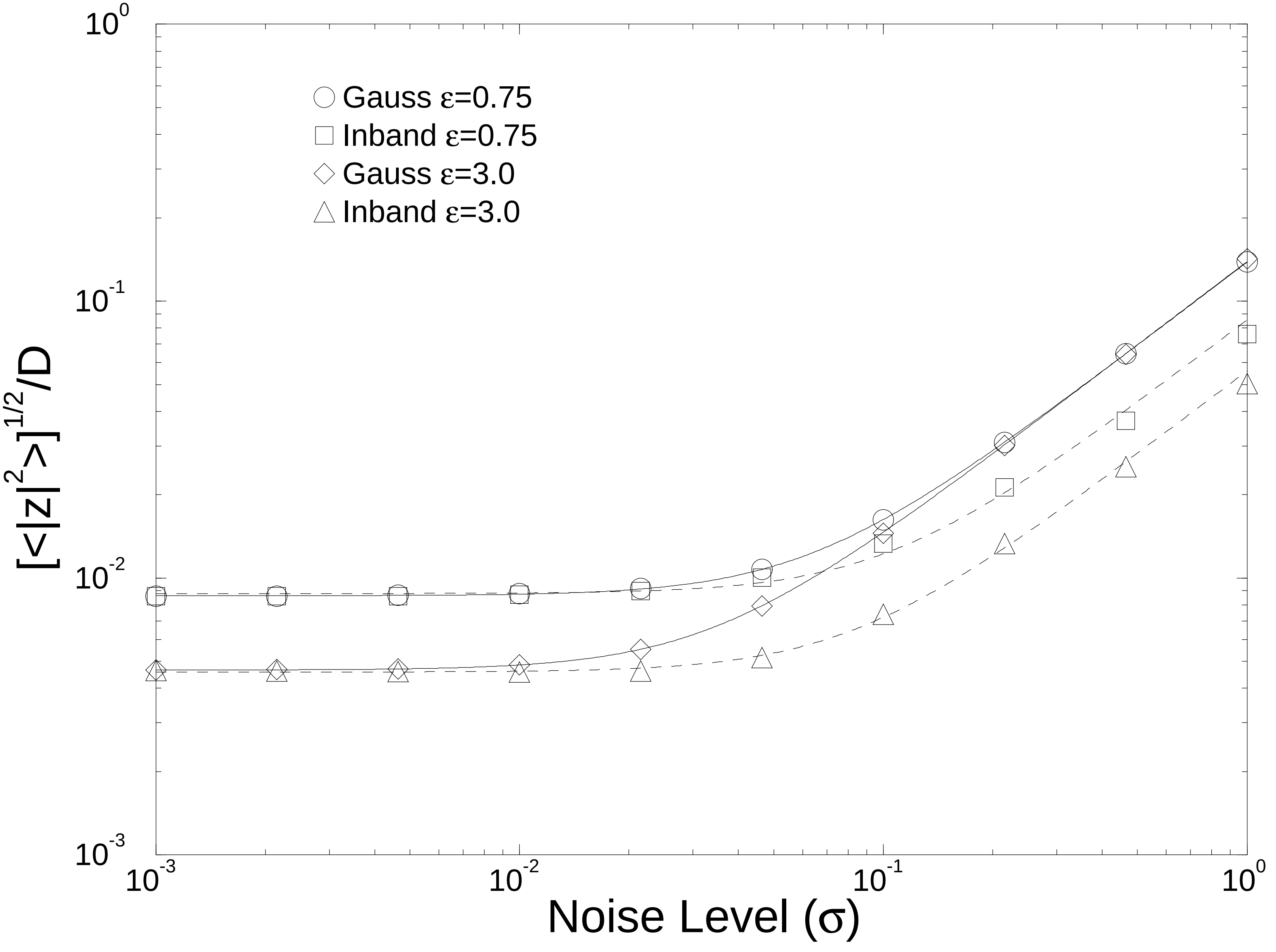

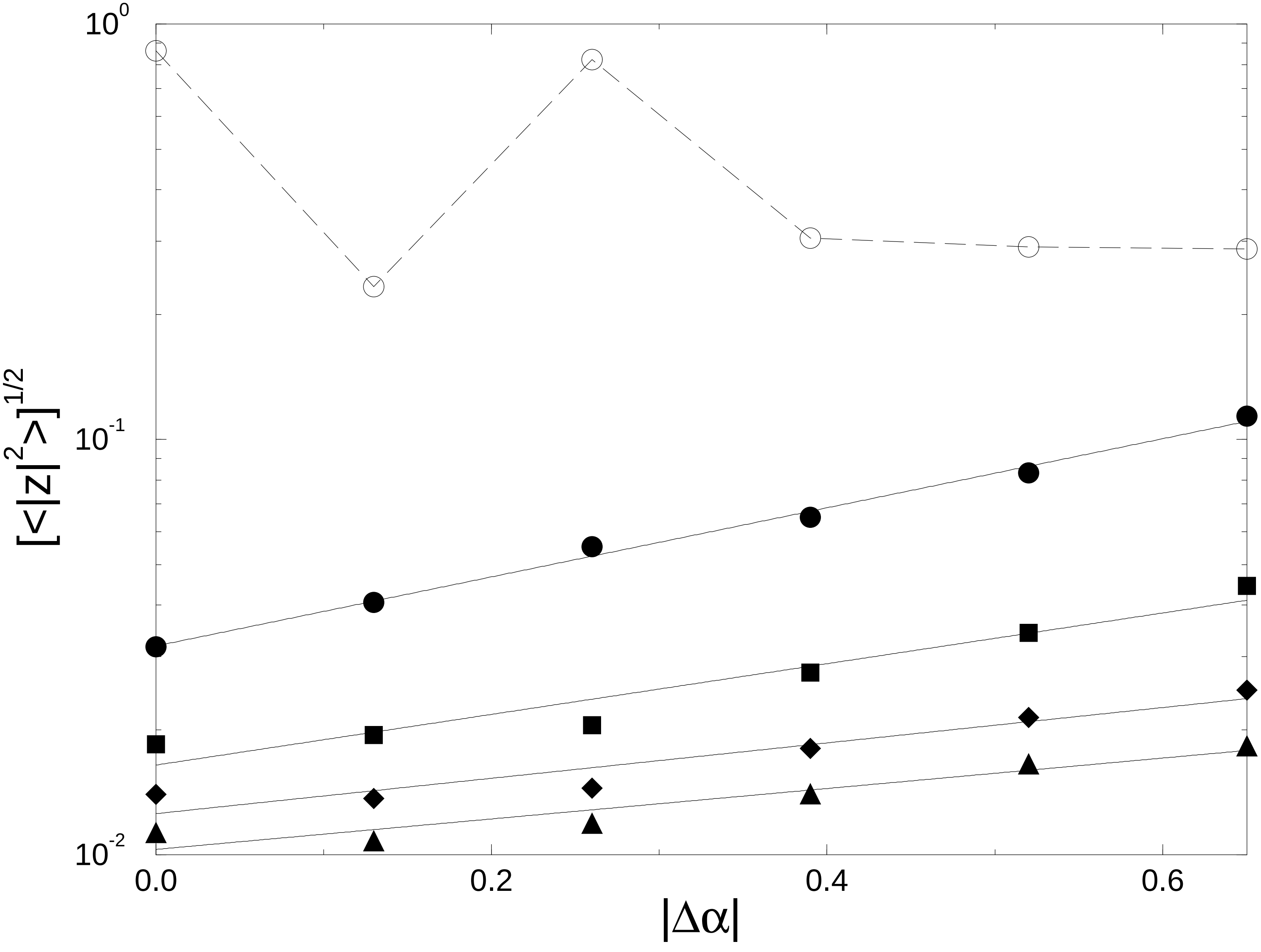

10^0
10^-1
10^-2
0.0
0.2
0.4
0.6
[<|z|^2>]^1/2
|Δα|

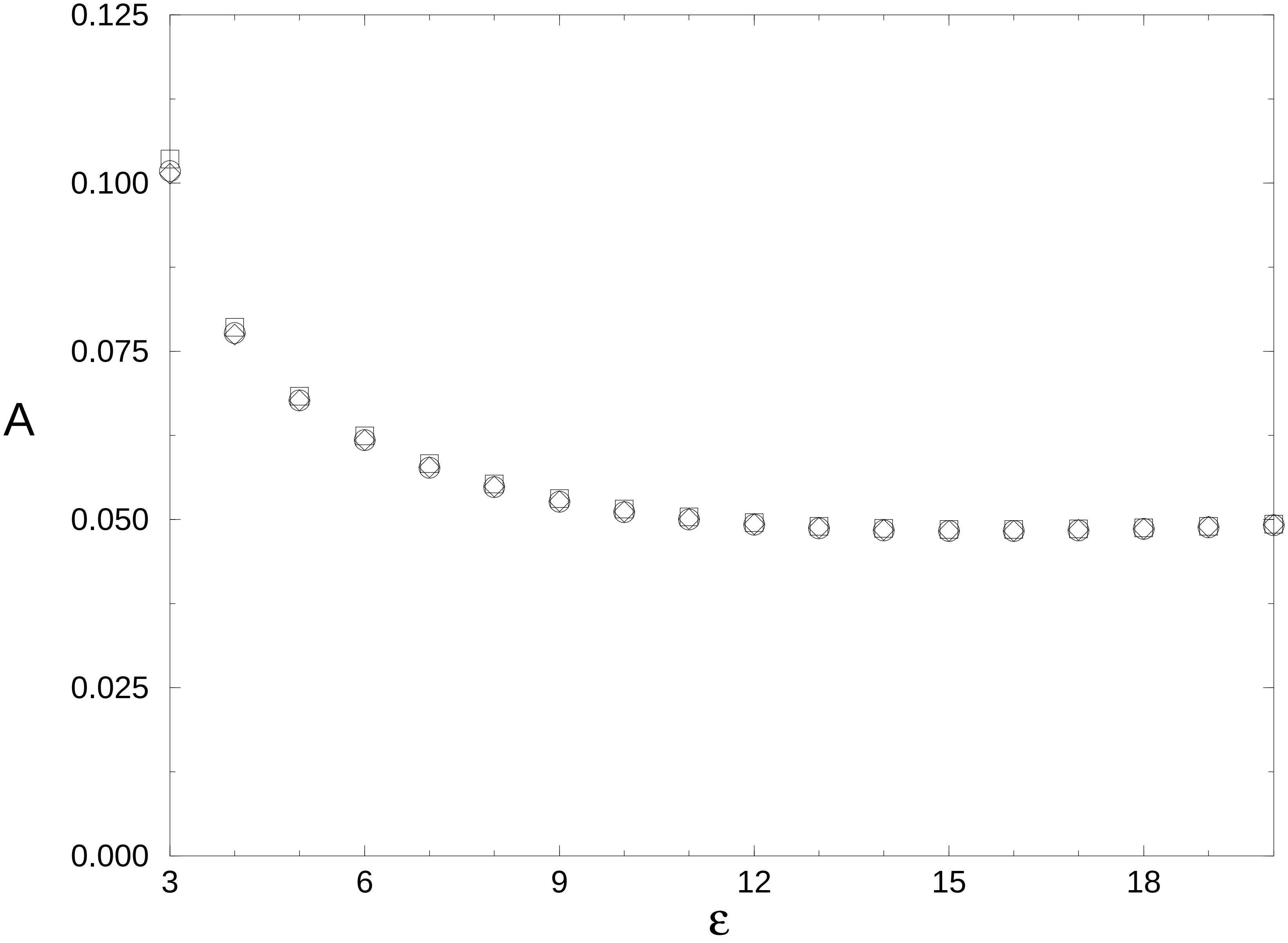

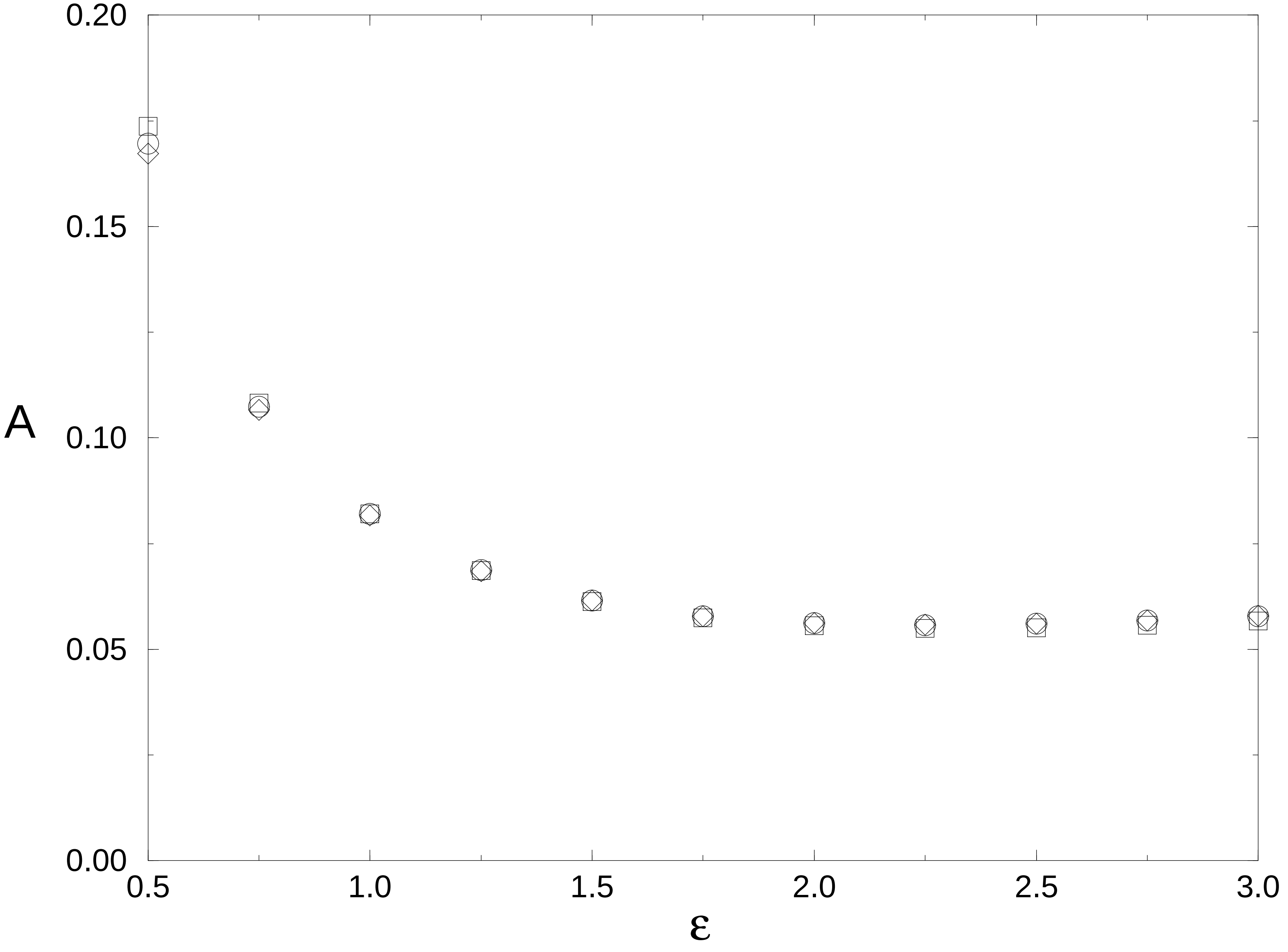

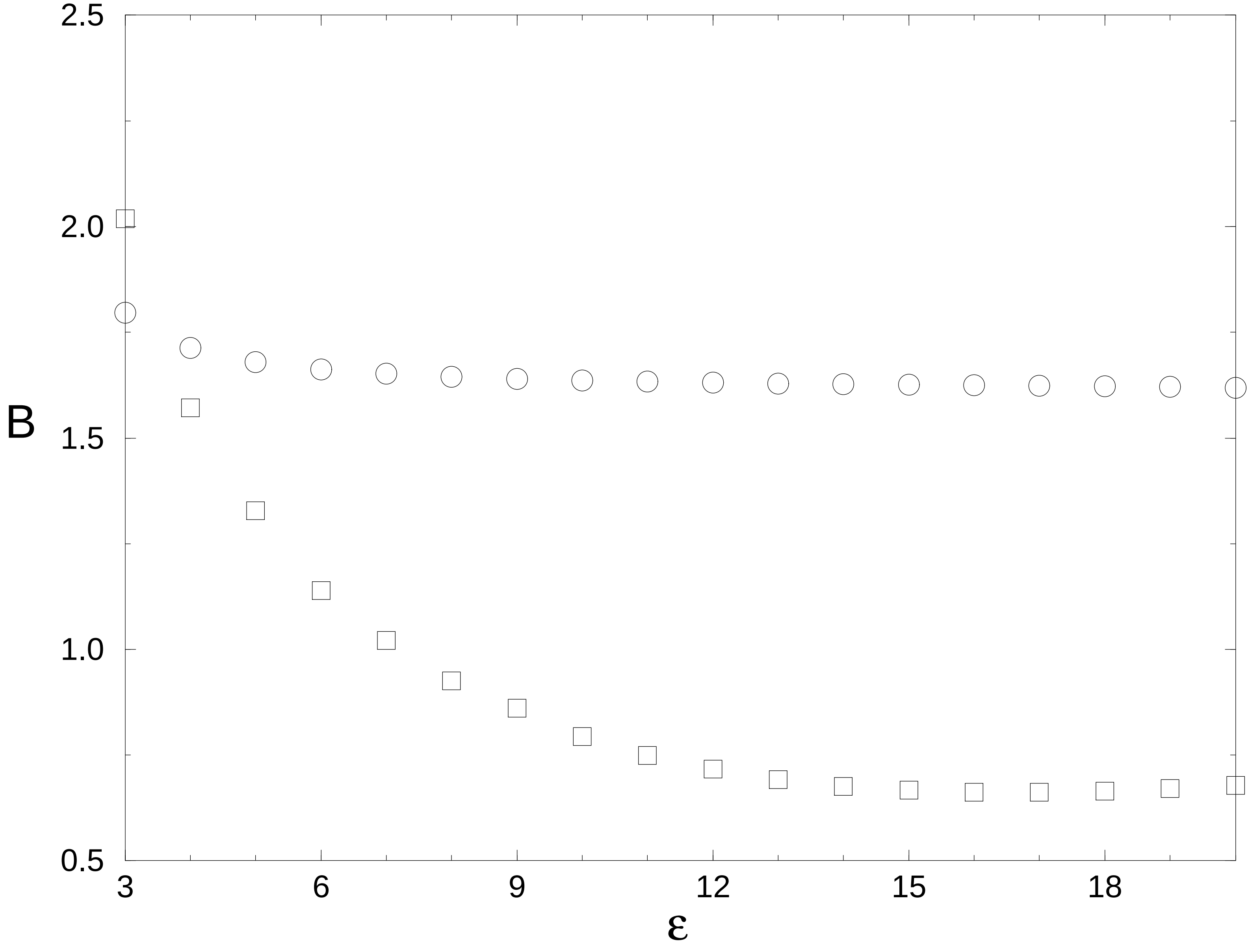

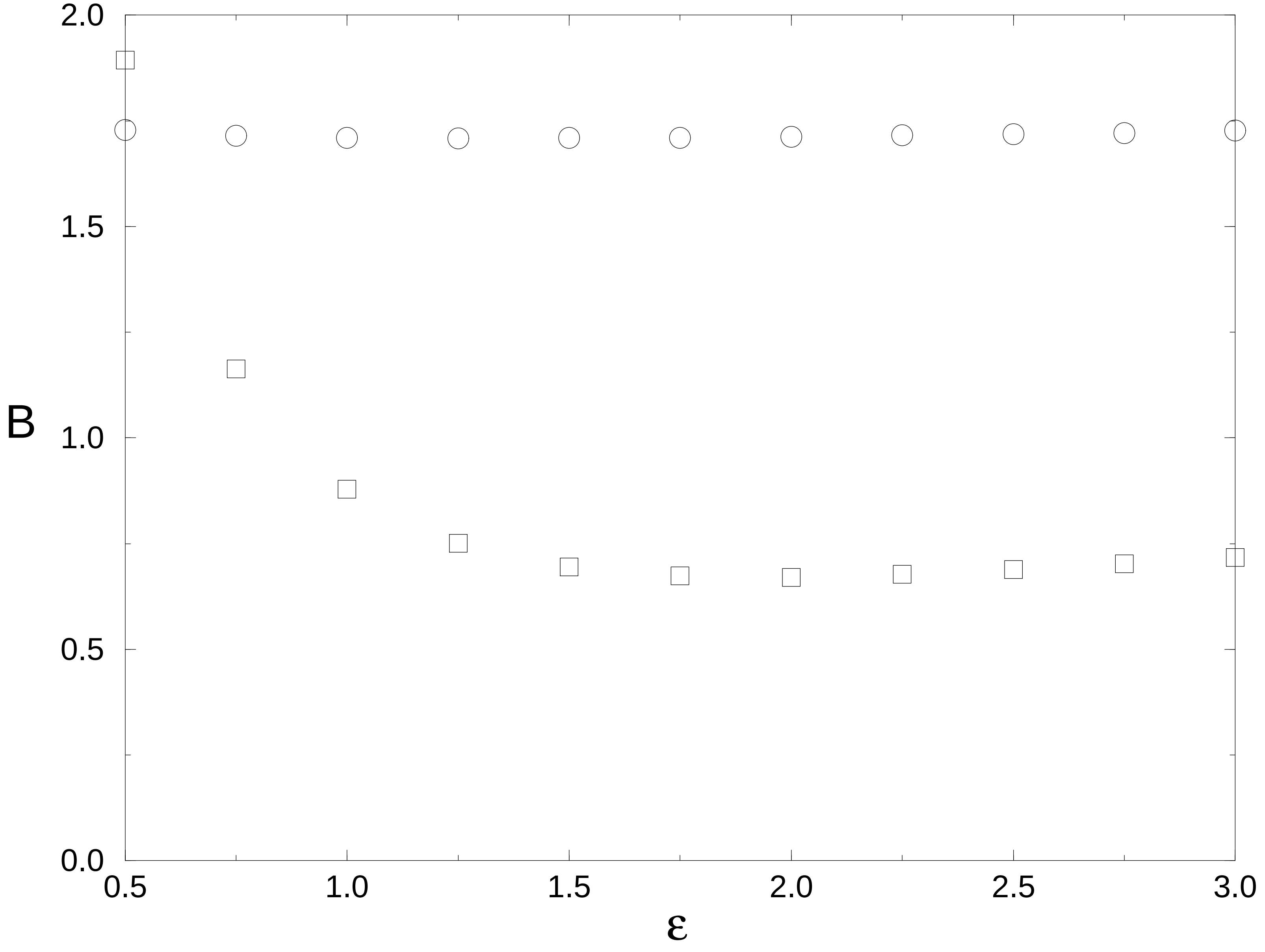

# The effects of additive noise and drift in the dynamics of the driving on chaotic synchronization

Reggie Brown and Nikolai F. Rulkov
*Institute for Nonlinear Science, University of California, San Diego, La Jolla, CA 92093-0402*

Nicholas B. Tufillaro
*Center for Nonlinear Studies, Los Alamos National Laboratory, Los Alamos, New Mexico 87545*

We examine the effect of additive noise and drift in the dynamics of a chaotic driving signal on the synchronization of a chaotic response systems. Simple scaling laws associated with the synchronization deviation level under these types of contamination are presented. Time series used as the driving signals are experimentally measured from an electronic circuit and a vibrating wire.

## I. INTRODUCTION

Synchronization between two chaotic systems has received considerable attention in recent years. In most of this work synchronization of chaos has been associated with identical, chaotic in time, behaviors brought about by coupling two or more identical systems in a drive/response manner [1,2]. (There are far too many papers on this subject to present a listing of all useful references. A partial listing can be found in the paper by Heagy, Carroll, and Pecora [3] or our longer paper [4]. The topic is also discussed in the review by Abarbanel et. al. [5].) Unless otherwise specified synchronization will mean this form of identical synchronous motion of identical systems.

This letter examines the following questions regarding the appearance of deviations from identical synchronous motion: (1) How will additive noise in the driving signal effect synchronization between the systems? (2) How will small differences between the dynamics of the driving and response systems effect synchronization between the systems? Similar questions have been addressed by others [2,6].

To date, the primary suggested application for synchronization involves communications [7–9]. All of the proposed communication methods that use a chaotic carrier to mask the signal of interest require synchronization (identical synchronous motion) between the transmitter and the receiver. Noise in the transmission channel, or difference in the dynamics, will introducing deviations from synchronization which are not part of the signal of interest. The two questions we address also have application when synchronization is used as a form of nondestructive testing, failure monitoring, and system identification [4].

Our research uses numerical models constructed from experimentally measured time series data as the response system [10]. Time series measurements are also used as the driving system. The "working phase space" is the one where the global dynamics is being modeled [5,11]. Typically, the working phase space is a $d$ dimensional Euclidean space reconstructed from time series measurements. Let the unknown dynamics of the driving system in the working phase space be represented by

$$\frac{d\mathbf{x}}{dt} = \mathbf{G}(\mathbf{x}). \tag{1}$$

The only thing known, *a priori*, about $\mathbf{G}$ are time series measurements. Let the dynamics of the model in the same phase space be represented by

$$\frac{d\mathbf{x}}{dt} = \mathbf{F}(\mathbf{x}). \tag{2}$$

Figure 1 is a pictorial example of an attempt to synchronize a model to a time series. To generate this picture a time series from an electronic circuit (see Section III) is used to construct a model, Eq. (2), of the dynamics of the circuit. Next, we recorded $\mathbf{x}$, a second trajectory from the circuit. $\mathbf{F}$ and $\mathbf{x}$ were used to generate new trajectories, $\mathbf{w}$ and $\mathbf{y}$. The trajectory, $\mathbf{w}$, results when the first point in $\mathbf{x}$ is used as an initial condition and the model, Eq. (2), is integrated forward in time without driving. The trajectory, $\mathbf{y}$ results when the model is driven by $\mathbf{x}$ (using the driving method discussed in Section II).

The lower curve in Fig. 1 is $|\mathbf{z}|^2 = |\mathbf{x} - \mathbf{y}|^2$. The upper curve is $|\mathbf{z}|^2 = |\mathbf{x} - \mathbf{w}|^2$. The fact that the distances shown in the lower curve are small, when compared to those in the upper curve, indicates that the deviations from identical synchronous motion are small.

In the complete absence of noise and modeling errors we expect $|\mathbf{z}|^2 = 0$, and the lower curve in Fig. 1 resides at $-\infty$. However, since noise is always present in the driving signal and there are always modeling errors, two physical devices can only be almost synchronized. As either the size of the noise or the size of the modeling errors increases the level of the lower curve in Fig. 1 rises. Understanding the behavior of this rise as a function of noise and modeling errors will answer questions (1) and (2).

## II. THEORETICAL CONSIDERATIONS

Assume that, in the working phase space, Eq. (1) represents the true dynamics of the device that is producing the driving signal. The driving signal, $\mathbf{x} + \sigma\hat{\mathbf{u}}$, is the sum of the clean dynamics, $\mathbf{x}$, and the noise, $\sigma\hat{\mathbf{u}}$. The size of the noise is given by $\sigma$ while $\hat{\mathbf{u}}$ is a signal of unit size. Now assume that a model, Eq. (2), has been constructed from a time series which is not the same as the driving time series (although they may have the same source) [10,12].

By dissipatively coupling $\mathbf{F}$ to the driving signal via

$$\frac{d\mathbf{y}}{dt} = \mathbf{F}(\mathbf{y}) - \mathbf{E} \cdot [\mathbf{y} - (\mathbf{x} + \sigma\hat{\mathbf{u}})], \tag{3}$$

it is possible to almost synchronize $\mathbf{F}$ to $\mathbf{x}$ [10,13]. The coupling matrix $\mathbf{E}$ has only one nonzero element, $E_{\beta\beta} = \epsilon$, where $x_\beta + \sigma\eta$ is the $\beta$ component of the driving signal. Synchronized, or almost synchronized, motion may only be possible within some finite range of $\epsilon$. Under these conditions using a value of $\epsilon$ that is too small *or* too large will not produce synchronous, or almost synchronous motion [2,10,13]. Thus, errors in modeling and noise are the causes of any lack of complete synchronization in Eq. (3). For our numerical experiments the $\epsilon \to \infty$ limit still results in near synchronization when $\beta$ is chosen to be $d$, the last component of the driving vector.

By definition, the model is synchronized to the time series, $\mathbf{x}$, if $\mathbf{x} = \mathbf{y}$ *and* $\mathbf{F} = \mathbf{G}$ for all time greater than $t_0$, a transient. It is easy to show that synchronization will not occur when $\mathbf{F} \neq \mathbf{G}$ and/or in the presence of noise [10]. Let $\mathbf{z} = \mathbf{y} - \mathbf{x}$ denote the deviations between $\mathbf{y}$ and the clean driving signal, $\mathbf{x}$. If the deviations are small, $0 < |\mathbf{z}|^2 \ll 1$, then the linearized time evolution of $\mathbf{z}$ is given by

$$\frac{d\mathbf{z}}{dt} = [\mathbf{DF}(\mathbf{x}) - \mathbf{E}] \cdot \mathbf{z} + \sigma\mathbf{E} \cdot \hat{\mathbf{u}} + \Delta\mathbf{G}(\mathbf{x}) \tag{4}$$

where $\Delta\mathbf{G} = \mathbf{F} - \mathbf{G}$ denotes the difference between the model and the true dynamics of the driving system.

$\Delta\mathbf{G}$ is not related to measurement errors and has two potential sources. The first source arises because for any real situation $\mathbf{F}$ is never *exactly* equal to $\mathbf{G}$. The second source arises if the dynamics of the driving signal, $\mathbf{G}$, is different from the dynamics that produced the time series used to make the model, $\mathbf{G}'$. To analytically isolate these two sources assume that $\mathbf{G}$ and $\mathbf{G}'$ are related by some small change in the parameters of the driving system, $\mathbf{G} \simeq \mathbf{G}' + (\partial\mathbf{G}'/\partial\mathbf{p}) \cdot \delta\mathbf{p}$. Then $\Delta\mathbf{G}$ in Eq. (4) becomes

$$\Delta\mathbf{G}(\mathbf{x}) \simeq \Delta\mathbf{G}'(\mathbf{x}) + \left(\frac{\partial}{\partial\mathbf{p}}\Delta\mathbf{G}'(\mathbf{x})\right) \cdot \delta\mathbf{p}, \tag{5}$$

and $\Delta\mathbf{G}' = \mathbf{F} - \mathbf{G}'$. The first and second terms on the right hand side of in this equation are associated with modeling errors and drift in the dynamics of the driving system, respectively.

Equations (4) and (5) are the principle evolution equations for coupled systems in the vicinity of synchronized chaotic motion. An expression for $\mathbf{z}(t)$ can be obtained by formally integrating Eq. (4) [4].

The bottom curve in Fig. 1 indicates that the average behavior of $\log_{10}(|\mathbf{z}|^2)$ is essentially constant. Therefore, we define the synchronization deviation level (which we sometimes call the deviation level) by

$$\left[\langle|\mathbf{z}|^2\rangle_T\right]^{1/2} = \left[\lim_{t\to\infty} \frac{1}{t - t_0} \int_{t_0}^{t} |\mathbf{z}(r)|^2 dr\right]^{1/2}. \tag{6}$$

At this time we make the following assumptions regarding the noise and the errors in modeling. We assume that the noise, and $\Delta\mathbf{G}$, are ergodic. This allows us to replace time averages by phase space averages (which will be denoted by brackets, $\langle\rangle$). We also assume that the noise is stationary and completely independent of $\Delta\mathbf{G}$. The stationary assumption implies that the autocorrelation of the noise, $k(s) = \langle\eta(r)\eta(r - s)\rangle$, is only a function of the lag time, $s$.

With these assumptions it can be shown that Eq. (6) can be rewritten as [4]

$$\left[\left\langle |\mathbf{z}|^2\right\rangle_T\right]^{1/2} = \left[A^2 + (\sigma B)^2\right]^{1/2}, \tag{7}$$

where $B^2$ and $A^2$ are given by

$$B^2 = \epsilon^2 k(0)\left[\int_{t_0}^{t}\left\langle |\mathbf{U}(t,r)\cdot\mathbf{V}|^2\right\rangle dr + 2\int_{t_0}^{t}\left\langle [\mathbf{U}(t,r)\cdot\mathbf{V}]\cdot[\mathbf{U}(t,r)\cdot\mathbf{B}(r)]\right\rangle dr\right], \tag{8}$$

and

$$A^2 = \int_{t_0}^{t}\left\langle |\mathbf{U}(t,r)\cdot\Delta\mathbf{G}(r)|^2\right\rangle dr + 2\int_{t_0}^{t}\left\langle [\mathbf{U}(t,r)\cdot\Delta\mathbf{G}(r)]\cdot[\mathbf{U}(t,r)\cdot\mathbf{H}(r)]\right\rangle dr. \tag{9}$$

The three $d$ dimensional vectors, $\mathbf{V}$, $\mathbf{B}(r) = \mathbf{B}[\mathbf{x}(r)]$ and $\mathbf{H}(r) = \mathbf{H}[\mathbf{x}(r)]$ are defined in the following manner. $\mathbf{V} = [0,\ldots,0,1,0,\ldots,0]$ where the 1 appears as the $\beta$ element of $\mathbf{V}$ and $\beta$ is given by $E_{\beta\beta} \neq 0$. $\mathbf{B}(r)$ and $\mathbf{H}(r)$ are defined by

$$\mathbf{B}(r) = \int_{0+}^{r}\frac{k(s)}{k(0)}\mathbf{U}(r,r-s)\cdot\mathbf{V}\,ds$$
$$\mathbf{H}(r) = \int_{0+}^{r}\mathbf{U}(r,r-s)\cdot\Delta\mathbf{G}(r-s)\,ds,$$

where $\Delta\mathbf{G}(r) = \Delta\mathbf{G}[\mathbf{x}(r)]$. The lower limit of integration, $0^{+}$, implies taking the limit as we approach, but are never equal to, 0 from the positive side.

The matrix $\mathbf{U}(t,t_0)$ satisfies the initial condition $\mathbf{U}(t_0,t_0) = \mathbf{1}$ (where $\mathbf{1}$ is the identity) and is the evolution operator that evolves the initial condition, $\mathbf{z}(t_0)$, forward in time from $t_0$ to $t$ in the absence of noise and modeling errors. It comes from the solution to the homogeneous portion of Eq. (4) and is defined by

$$\begin{aligned}\mathbf{z}(t) &= \exp\left[\int_{t_0}^{t}[\mathbf{DF}(r)-\mathbf{E}]\,dr\right]\cdot\mathbf{z}(t_0)\\ &= \mathbf{U}(t,t_0)\cdot\mathbf{z}(t_0),\end{aligned}$$

where $\mathbf{DF}(r) = \mathbf{DF}[\mathbf{x}(r)]$. Since the time evolution is stable (the system synchronizes) $\mathbf{U}(t,t_0)$ shrinks $\mathbf{z}(t_0)$ to zero exponentially fast as $t \to \infty$. The rate of decrease is controlled, in a nontrivial fashion, by the coupling, $\epsilon$ [3,14].

Clearly, $B$ and $A$ are nontrivial functions of $\epsilon$. In addition, $B$ is a function of the type of noise in the driving signal but is *not* a function of $\sigma$ or the errors in modeling. On the other hand, $A$ is a function of modeling errors but is *not* a function of the noise. The scaling law given by Eq. (7) describes the rise of the deviation level as the noise level increases.

The second question asked in the introduction involved modeling errors and the effects of drift in the dynamics of the driving signal. Both of these effects influence $A$ while neither influences $B$. Inserting Eq. (5) and the definition of $\mathbf{H}(r)$ into Eq. (9) results in

$$A^2 \simeq A'^2 + \frac{\partial}{\partial\mathbf{p}}\left[\int_{t_0}^{t}\left\langle |\mathbf{U}(t,r)\cdot\left(\Delta\mathbf{G}'(r)+\mathbf{H}'(r)\right)|^2\right\rangle dr\right]\cdot\delta\mathbf{p}, \tag{10}$$

where $\mathbf{H}'$ is what one obtains by substituting $\Delta\mathbf{G}'$ for $\Delta\mathbf{G}$ in the definition of $\mathbf{H}$, and $A'^2$ is what one obtains by substituting $\Delta\mathbf{G}'$ and $\mathbf{H}'$ for $\Delta\mathbf{G}$ and $\mathbf{H}$ in Eq. (9).

In the limit of zero noise the deviation level is just $\left[\left\langle |\mathbf{z}|^2\right\rangle_T\right]^{1/2} = A$. Thus, in this limit Eq. (10) describes the rise of the deviation level as the dynamics of the driving drifts. The ability to track this rise is the center of a nondestructive testing application we discuss in our longer paper [4].

## III. NUMERICAL EXPERIMENTS

In Section II we addressed changes in the synchronization deviation level as a function of noise level, $\sigma$, and modeling errors, $\Delta\mathbf{G}$. Our numerical experiments examine each of these issues separately and are performed on data sets taken from an electronic circuit and a vibrating wire.

Details about the experimental apparatus that produced the scalar time series, $s(n\Delta t) = s(n)$, $n = 1, 2, \ldots$ can be found in our longer paper and references therein [4]. The time delay method

$$\mathbf{x}(n) = [s(n),\, s(n+T), \ldots, s(n+(d-1)T)]$$

is used to reconstruct the attractors in a working phase space. The optimal time delays and embedding dimensions were determined using average mutual information and false near neighbors, respectively [4,15,16]. The experimental systems we have examined produced time series data that was embeddable in $d = 3$ dimensions.

Next we used portions of the embedded time series to constructed global ODE's in the form of Eq. (2) to serve as models for the dynamics on the attractors [4,10]. All of the numerical experiments used the third component of the embedded time series as the driving term (i.e., $\beta = d = 3$ in Eq. (3).

In our first set of numerical experiments we examined the behavior of $[\langle |\mathbf{z}|^2 \rangle_T]^{1/2}$ as a function of $\sigma$. Each numerical experiment used two types of noise. The first type is gaussian noise with zero mean and unit standard deviation. The second type (inband noise) is constructed to have zero mean, unit standard deviation, and the same power spectrum as the raw time series. We also investigated two different values of the coupling constant, $\epsilon$. One value was chosen slightly above the minimum necessary for synchronization while the second was chosen well above the minimum value.

The numerical tests used 5000 and 10,000 point time averages for the circuit and wire, respectively. The results are shown in Figs. 2 where we have plotted normalized deviation levels, $[\langle |\mathbf{z}|^2 \rangle_T]^{1/2}/D$. The normalization constant, $D$, is associated with the time average of $|\mathbf{z}|^2$ when $\epsilon = 0$. The lines represent curves of best fit between Eq. (7) and the results of the numerical experiments.

A second set of numerical experiments we performed involved determining the behavior of the deviation level, $[\langle |\mathbf{z}|^2 \rangle_T]^{1/2}$, as a function of changes in the dynamics of the driving signal, $\Delta\mathbf{G}$. In order to perform our numerical tests we recorded six time series from the circuit each corresponding to a slightly different value of a parameter, $\alpha_0, \ldots \alpha_5$, (physically, a resistance is changed in the circuit).

A model, Eq. (2), was constructed from a portion of the $\alpha_0$ time series and then subjected to driving from each time series. The measured deviation levels are shown in Fig. 3, where the solid lines represent straight lines of best fit through the data. The solid circles, squares, diamonds, and triangles are associated with coupling strengths of $\epsilon = 1$, 2, 3, and 4, respectively. The open circles represent a test case with a coupling strength of $\epsilon = 0.5$ which is insufficient to synchronize the model to the driving signal. The solid symbols indicate a linear rise in the deviation level with respect to changes in $\alpha$ ($|\Delta\alpha_j| = |\alpha_0 - \alpha_j|$).

The rise in the deviation level shown in Fig. 3 is due, predominantly, to changes in the dynamics of the driving signal [4]. If the amplitude of the noise is small then Eqs. (7), (9), and (10) imply that

$$\log_{10}\left( [\langle |\mathbf{z}|^2 \rangle]^{1/2} \right) \simeq \log_{10}(A') + \frac{1}{2}\left(\frac{S}{A'}\right)^2 |\Delta\alpha|,$$

for suitably defined $S(\epsilon)$ (see Eq. (10)). This equation indicates that the deviation level should rise linearly with $|\Delta\alpha|$ as shown in Fig. 3.

Figures 4 show $A$ vs $\epsilon$ for three distinct cases. The first case (diamonds) used the raw data as the driving term in Eq. (3) and the approximation $A^2 = \langle |\mathbf{z}|^2 \rangle_T$. The second two cases (circles and squares) used gaussian and inband noise, respectively. (Figure 4a shows results from the electronic circuit while Fig. 4b shows results from the vibrating wire.) To obtain $A$ (as well as $B$) for a particular value of $\epsilon$ we calculated the deviation level as a function of noise size (see Fig. 2). We then fitted these results to Eq. (7) [18]. The figures indicates that $A$ is not a function of the type of noise in the driving.

Figures 5 shows $B$ vs $\epsilon$ for inband and gaussian noise. (Figure 5a shows results from the electronic circut while Fig. 5b shows results from the vibrating wire.) Quantitative details about the behavior of $B$ depend intimately on the type of noise that is found in the driving signal, and will change from one type of noise to the next. An important special case is delta correlated noise where $k(s) = k(0)\,\delta(s)$. When the noise is delta correlated $\mathbf{B} = \mathbf{0}$ and it is possible to obtain a compact analytic expression for $B^2$ [4]. The figures indicate that $B$ is essentially independent of $\epsilon$ for gaussian noise and has a strong dependence on $\epsilon$ for inband noise.

Figures 4 and 5 show that both $B$ and $A$ appears to become independent of $\epsilon$ when $\epsilon$ gets large. This fact can be predicted since in the $\epsilon \to \infty$ limit corresponds to Pecora and Carroll synchronization [4,1]. A partial theoretical analysis of the dependence of $B$ and $A$ on $\epsilon$ can be found in our longer paper [4].

## IV. SUMMARY AND CONCLUSIONS

This letter has investigated the behavior of chaotic synchronization in the presence of additive noise in the driving and drift in the dynamics of the driving. We defined a measure of the deviation from synchronization between the driving signal and the response system. If the driving signal is denoted by $\mathbf{x}$ and the dynamics of the response system while being driven is denoted by $\mathbf{y}$, then the synchronization deviation level is given by the following time average $\left[\langle|\mathbf{z}|^2\rangle_T\right]^{1/2}$ where $\mathbf{z} = \mathbf{y} - \mathbf{x}$ (see Section II).

The numerical experiments involved constructing models of the dynamics of the driving system from experimentally measured time series. These models are the response systems which were driven by time series data.

To study the effect of additive noise in the driving signal on $\left[\langle|\mathbf{z}|^2\rangle_T\right]^{1/2}$ we added gaussian as well as inband noise of various amplitudes to the driving signal. Theoretical analysis in Section II indicates that $\langle|\mathbf{z}|^2\rangle_T = A^2 + (\sigma B)^2$ where $\sigma$ is the standard deviation of the noise in the driving signal, and $B$ and $A$ are given by Eqs. (8) and (9). The results of our numerical experiments and confirmed this scaling law.

To study the effects of drift in the dynamics of the driving signal on $\left[\langle|\mathbf{z}|^2\rangle_T\right]^{1/2}$ we constructed a model using data from the electronic circuit. The model was then subjected to driving from time series obtained from the circuit *after* making slight changes in one of its components (a resistor). The results are shown in Fig 3 and confirm the theoretical analysis shown in Sections II and III.

The quantities $B$ and $A$ are functions of the coupling strength, $\epsilon$. We have presented figures displaying the dependence of $B$ and $A$ on $\epsilon$ for both of the systems studied. A more complete analysis can be found elsewhere [4].

## V. ACKNOWLEDGMENTS

N. Rulkov was supported by grant number DE-FG03-90ER14148 from the Department of Energy. N. Tufillaro was supported by the Department of Energy.

FIG. 1. The distances between the orbits $\mathbf{x}$, $\mathbf{y}$, and $\mathbf{w}$. The dashed line is $\mathbf{z} = |\mathbf{x} - \mathbf{y}|$ while the solid line is $|\mathbf{x} - \mathbf{w}|^2$. Near synchronization between $\mathbf{x}$ and $\mathbf{y}$ is clearly demonstrated. The driving variable is $x_3$ and $\epsilon = 20$.

FIG. 2. The normalized synchronization levels as a function of added noise level. The circles and squares correspond to gaussian and inband noise for the smaller value of $\epsilon$, respectively. The diamonds and triangles correspond to gaussian and inband noise for larger value of $\epsilon$, respectively. For the circuit we used $\epsilon = 5$ and 20 while for the wire we used $\epsilon = 0.75$ and 3.

FIG. 3. The synchronization level as a function of $\Delta\alpha$ and the coupling strength, $\epsilon$. Each change in $\Delta\alpha$ represents an change in the dynamics of the driving signal of approximately 1 %.

FIG. 4. $A$ vs $\epsilon$. $A$ has been calculated using raw data (the diamonds) data with gaussian noise (the circles) and data with inband noise (the squares). (a) Results for the electronic circuit. (b) Results for the vibrating wire.

FIG. 5. A graph of $B$ vs $\epsilon$ for gaussian (the circles) and inband noise (the squares). (a) Results for the electronic circuit. (b) Results for the vibrating wire.